# Observation of nonlinearity-controlled switching of topological edge states

A. A. Arkhipova,[1,2] S. K. Ivanov,[1] S. A. Zhuravitskii,[1,3] N. N. Skryabin,[1,3] I. V. Dyakonov,[3] A. A. Kalinkin,[1,3] S. P. Kulik,[3] V. O. Kompanets,[1] S. V. Chekalin,[1] Y. V. Kartashov,[1] and V. N. Zadkov[1,2]

[1]*Institute of Spectroscopy, Russian Academy of Sciences, 108840, Troitsk, Moscow, Russia*
[2]*Faculty of Physics, Higher School of Economics, 105066 Moscow, Russia*
[3]*Quantum Technology Centre, Faculty of Physics, M. V. Lomonosov Moscow State University, 119991, Moscow, Russia*
*Corresponding author: Yaroslav.Kartashov@icfo.eu

We report the experimental observation of the periodic switching of topological edge states between two dimerized fs-laser written waveguide arrays. Switching occurs due to the overlap of the modal fields of the edge states from topological forbidden gap, when they are simultaneously present in two arrays brought into close proximity. We found that the phenomenon occurs for both strongly and weakly localized edge states and that switching rate increases with decreasing spacing between the topological arrays. When topological arrays are brought in contact with nontopological ones, switching in topological gap does not occur, while one observes either the formation of nearly stationary topological interface mode or strongly asymmetric diffraction into the nontopological array depending on the position of the initial excitation. Switching between topological arrays can be controlled and even completely arrested by increasing the peak power of the input signal, as we observed with different array spacings.

Topological insulators are of salient importance in several areas of physics and they have been realized in multiple systems. Their importance is motivated by the rich propagation properties and exceptional robustness of the edge excitations that appear at the interfaces between two materials described by Hamiltonians characterized by different topological invariants. In such systems, excitations can propagate along the edges even when the bulk is insulating, and they are fundamentally protected by topology, such as, for example, in systems that exhibit band structures with nontrivial topology. The phenomenon of topological insulation, first predicted in solid-state physics [1,2], have been since then studied theoretically and realized experimentally in mechanical systems [3,4], acoustics [5-7], atomic systems in optical lattices [8-12], with exciton-polaritons in structured microcavities [13-20], and in photonic [21-26] and many other systems. The recent progress in the subfield of topological photonics, where such insulators may find direct practical applications in addition to their fundamental physics interest, is presented in the reviews [27-30].

The topological protection of edge states makes them ideal candidates for the realization of various switching and routing schemes, and for the exploration of photonic circuits based on controllable disorder-resistant transfer of topological states across the circuit [31-33]. Investigation of such systems and coupling mechanisms is a task of particular relevance due to their potential ultrafast response. To date, various coupling mechanisms for topological edge states have been suggested. They include topological pumping schemes involving the transfer of excitations between the opposite edges of the topologically nontrivial structures [34,35], utilization of gradients along the topological insulator edge leading to switching of excitations between opposite edges due to anomalous Bloch oscillations [36,37], shallow resonant modulations of the underlying potential landscapes leading to Rabi oscillations between the edge states [38,39], switching controlled by the resonant pump in hollow and ribbon-shaped microresonator arrays [40,41], coupling between Floquet edge states in helical waveguides arrays with opposite helicities [42] and coupling between states existing at the opposite edges of small-scale Su-Schrieffer-Heeger chains [43,44].

However, most of the above coupling schemes have been studied in purely linear regime, while driving topological system into the nonlinear regime may qualitatively modify the evolution dynamics and the conditions of formation of the edge states [45,46], allowing to tune their energies [47], triggering their efficient parametric interactions [48-50] and modulational instabilities [16,51,52], whose presence is also a direct indication of the fact that such systems may support solitons of topological origin in the bulk [53,54] or at the edge [55-65] of the insulator. Edge solitons have been observed in various topological photonic systems, including Floquet [62] and higher-order insulators [66,67]. In addition, nonlinearity can drive a topologically trivial system to a regime where topological phases emerge [68-71]; it also stimulates coupling between the topological states, similar to the coupling of higher-order corner modes via edge states reported in [72]. Nevertheless, the impact of nonlinearity on the coupling of edge states has not been observed experimentally yet.

In this work we report the experimental observation of nonlinearity-controlled switching between the edge states arising in two dimerized Su-Schrieffer-Heeger [73] topological waveguide arrays brought into close proximity. Each array, when properly truncated, can support solitons bifurcating from edge states in the topological bandgap, as realized theoretically [74-79] and experimentally [80-83]. Here we observe experimentally that when two of such arrays approach each other, the overlap of the modal fields for the topological states causes their periodic switching between the arrays, with a switching rate that depends on the intensity of the input beam and on the separation between the arrays. Importantly, we show that, in contrast to conventional couplers [84], such switching occurs for modes with staggered tails from the topological bandgap, whose localization strongly depends on the width of the topological gap. Increasing the nonlinearity can lead not only to the arrest of the coupling between the two arrays, but it may also cause an abrupt increase of the radiation into the bulk of the arrays.

For the theoretical analysis we employ the dimensionless continuous Schrödinger-type equation for the amplitude of the light field $\psi$ propagating in a medium with shallow transverse refractive index modulation that defines two spatially separated dimerized Su-Schrieffer-Heeger (SSH) arrays as follows:

$$i\frac{\partial \psi}{\partial z} = -\frac{1}{2}\left(\frac{\partial^2 \psi}{\partial x^2}+\frac{\partial^2 \psi}{\partial y^2}\right)-[\mathcal{R}_l(x,y)-\mathcal{R}_r(x,y)]\psi-|\psi|^2\psi, \quad (1)$$

Here the functions $\mathcal{R}_{l,r}(x,y)=p\sum_{m=1,2N}\mathcal{Q}(x-x_{lm,rm},y)$ describe the profiles of two spatially separated "line" arrays, $p$ is the refractive index modulation depth in each array and $x_{lm}, x_{rm}$ set the coordinates of the waveguide centers in the left, $l$, and right, $r$, arrays. Each array consists of $N$ pairs of waveguides (dimers) with identical Gaussian profiles $\mathcal{Q}(x,y)=e^{-(x^2+y^2)/a^2}$ of width $a$. Although the arrays considered here are effectively 1D line structures, we use a 2D model (1) to account for all possible instabilities in the focusing nonlinear medium. In the dimensionless Eq. (1), the $x,y$ coordinates are normalized to the characteristic transverse scale of $r_0 = 10\ \mu\mathrm{m}$, propagation distance $z$ to the diffraction length $kr_0^2$, while depth of the waveguides $p=k^2r_0^2\delta n/n$ is defined by the actual refractive index contrast $\delta n$. Here $k=2\pi n/\lambda$ is the wavenumber, $n\approx 1.45$ is the background refractive index. In these normalized units the waveguide width is $a=0.5$. Due to slight differences in writing conditions from day to day, we determine a specific value of $p$ using a test array written at the beginning of each array series – in our case $p\sim 4-5$, as indicated in figure captions for each particular case.

Microscope images of the fabricated arrays are presented in Fig. 1(a). Such arrays were inscribed in 10 cm-long fused silica glass samples by focused (with an aspheric lens with $\mathrm{NA}=0.3$) femtosecond laser pulses (wavelength $515\ \mathrm{nm}$, pulse duration $280\ \mathrm{fs}$, pulse energy $360\ \mathrm{nJ}$, repetition rate $1\ \mathrm{MHz}$). During the inscription process the sample was translated relative to the focus at a constant velocity of $1\ \mathrm{mm/s}$ using high-precision air-bearing positioner (Aerotech), resulting in the inscription of sets of parallel waveguides with controllable spacing between them. The depth of the refractive index modulation in such waveguides is about $\delta n\sim 5.5\times 10^{-4}$, i.e. they are single-mode with mode field diameter $\sim 15.4\times 24.0\ \mu\mathrm{m}$. Such waveguides exhibit about $0.3\ \mathrm{dB/cm}$ propagation losses at the wavelength of $800\ \mathrm{nm}$ used in experiments.

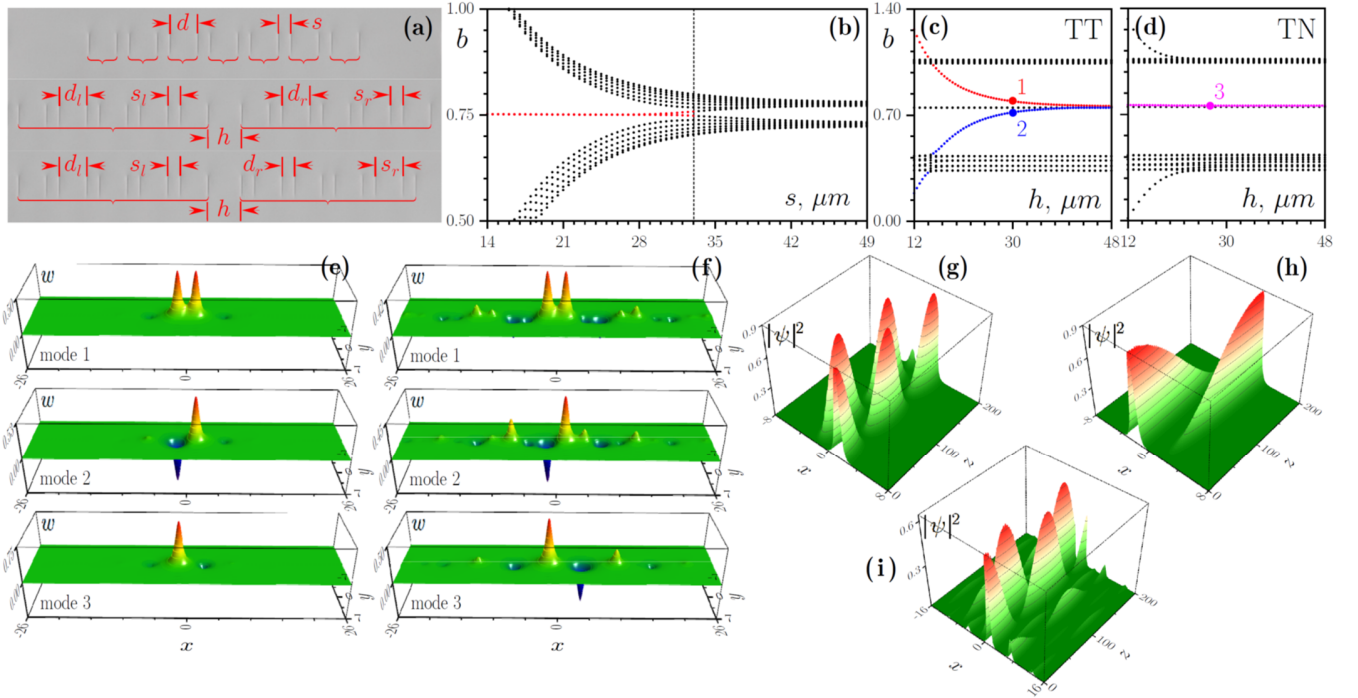

Fig. 1. (a) Photograph of the fs-laser written array of dimers in topological regime when inter-cell spacing $s=15\ \mu\mathrm{m}$ is smaller than intra-cell spacing $d=33\ \mu\mathrm{m}$ in the entire array (top). Photographs of the interface between two topological arrays with $s_l=s_r=15\ \mu\mathrm{m}$ and $d_l=d_r=33\ \mu\mathrm{m}$ (middle) and interface between topological and nontopological arrays with $s_l=d_r=15\ \mu\mathrm{m}$ and $s_r=d_l=33\ \mu\mathrm{m}$ (bottom). (b) Spectrum of modes of the SSH array with 7 dimers versus inter-cell spacing $s$ for fixed $d=33\ \mu\mathrm{m}$. Vertical dashed line indicates transition from the nontopological to the topological phase. Eigenvalues of modes supported by the topological-topological ($s_l=s_r=15\ \mu\mathrm{m}$, $d_l=d_r=33\ \mu\mathrm{m}$) (c) and topological-nontopological ($s_l=d_r=15\ \mu\mathrm{m}$, $s_r=d_l=33\ \mu\mathrm{m}$) (d) arrays as functions of spacing $h$ between two arrays. (e) Examples of the modes corresponding to dots in (c), (d). (f) Similar modes, but for $s_l=s_r=26\ \mu\mathrm{m}$, $d_l=d_r=33\ \mu\mathrm{m}$ (TT structure) and $s_l=d_r=26\ \mu\mathrm{m}$, $s_r=d_l=33\ \mu\mathrm{m}$ (TN structure) cases. Switching dynamics in TT structures for the same set of parameters as in (e) at $h=30\ \mu\mathrm{m}$ (g) and $h=40\ \mu\mathrm{m}$ (h), and for the parameters of (f) at $h=30\ \mu\mathrm{m}$ (i). In all cases, $p=4.88$.

In the SSH array a topological phase can be introduced by the opposite shifts of two waveguides within each unit cell (for a zero shift, the array becomes a usual periodic structure with identical separation between all the waveguides) [73]. When the inter-cell waveguide spacing $s$ becomes smaller than the intra-cell spacing $d$ due to such shift [as illustrated in the microscope image of single SSH array in the top row of Fig. 1(a)], the inter-cell coupling becomes stronger than the intra-cell one and the array enters into the topological phase. In this regime, in the spectrum of linear eigenmodes $\psi(x,y,z)=w(x,y)e^{ibz}$ [here $b$ is the propagation constant of the mode and $w(x,y)$ is the real function describing its profile] of the truncated array, one observes the appearance of the topological edge states (red dots), whose localization increases with decreasing spacing $s$ for fixed $d=33\ \mu\mathrm{m}$, see Fig. 1(b) for an exemplary spectrum of a *single* SSH array with $N=7$ dimers. The topological gap width, where edge states appear, also increases with decreasing $s$. At $s>d$ the array is in the trivial phase and no edge states can be formed in the trivial gap in the spectrum.

To realize coupling between the edge states we inscribed *two* SSH arrays, each with $N=5$ dimers, that are located in close proximity. Microscope images of two different configurations that we will consider below are presented in the middle and bottom rows of Fig. 1(a), where different arrays are highlighted by the horizontal red brackets. In the first TT configuration depicted in the middle row, both left and right arrays are in the topological phase, since inter-cell spacing $s_l = s_r = 15$ μm (here lower index denotes the array) in both arrays is smaller than the intra-cell one, $d_l = d_r = 33$ μm. The lower row in Fig. 1(a) shows other, TN configuration, when the left array is in the topological phase, while the right one is in the trivial phase that is achieved, for example, when $s_l = d_r = 15$ μm and $s_r = d_l = 33$ μm (i.e. the $s_r$ and $d_r$ values were swapped for the right array). The arrays are separated by a distance $h$ that we vary in the experiments.

To elucidate the possibility of switching of edge states, we calculated the linear spectrum of the TT configuration as a function of the spacing $h$ between the two topological arrays, see Fig. 1(c). One can clearly see a topological gap between the two bulk bands. The states corresponding to the red and blue lines within the topological gap are associated with in-phase and out-of-phase modes residing at the interface between the two arrays and penetrating into both of them [see modes 1 and 2, respectively, in Fig. 1(e)]. The propagation constants of such modes vary with the spacing $h$. The two modes within the gap with propagation constants that do not change with $h$ (black dots in the middle of the gap) are located at the outer edges of the entire structure and are not excited here. Excitation of the edge state only on the left or right side of the interface is equivalent to the simultaneous excitation of the in-phase mode 1 and out-of-phase mode 2 with nearly equal weights. Subsequent switching can be interpreted as a periodic beating between the interface modes, with a beating length that is inversely proportional to the propagation constant difference $L = \pi / (b_1 - b_2)$. $L$ increases with spacing $h$ due to the reduction of propagation constant difference. Examples of switching dynamics of the topological edge states for different spacing $h$ are presented in Fig. 1(g) and 1(h). Notice that we consider here arrays with sufficiently small $s_{l,r}$ values, which guarantee a strong localization of the edge states and their efficient excitation in the experiments. However, it should be stressed that switching takes place also for the weakly localized edge states, when $s_{l,r} \to d_{l,r}$. Examples of the interface states in the TT structure deeply penetrating into each array for these parameters are presented in Fig. 1(f). One clearly sees the staggered structure of the mode tails – an indication that such modes originate from the topological gap. The dynamics of switching for the weakly localized edge state is presented in Fig. 1(i).

In contrast to the above, when topological and nontopological arrays are brought in close proximity (TN structure), the eigenmode spectrum reveals in the topological gap the presence of only one localized mode 3 located near the interface [see magenta dots in Fig. 1(d), while black dots in the middle of the gap correspond to the edge state at the left outermost edge of the topological array that is not excited]. Such TN arrays are formally analogous to the arrays joined by a topological defect considered in [81,82], although in these works the spacing between defect waveguide and left and right nearest neighbors was fixed by the parameters of the unit cell, while here we deliberately vary spacing $h$ to see whether it can lead to the appearance of the new states that can affect switching dynamics. Mode 3 is mostly localized at the right outermost waveguide of the topological array [Fig. 1(e) and 1(f), last row], even though it can penetrate into nontopological array. Thus, in the TN structure the excitation at the left side of the interface will only generate localized mode 3, while excitation at the right edge should lead to enhanced diffraction, since no localized states exist in this part of the structure. In both cases no switching for modes from topological gap between arrays is predicted to occur. Notice that for sufficiently small $h$ values the spectrum of Fig. 1(d) reveals bifurcation of two modes from the top of the first band and bottom of the second band, whose propagation constants shift deeper into *nontopological* gaps with decrease of $h$. These localized modes with three in-phase or out-of-phase spots form at three closely spaced waveguides (right waveguide of the topological array and two left waveguides of the nontopological one), so that single-channel excitation leads to fast beatings between them. These modes, however, disappear (become delocalized) for sufficiently large spacing $h$ and we do not consider their switching because they are nontopological.

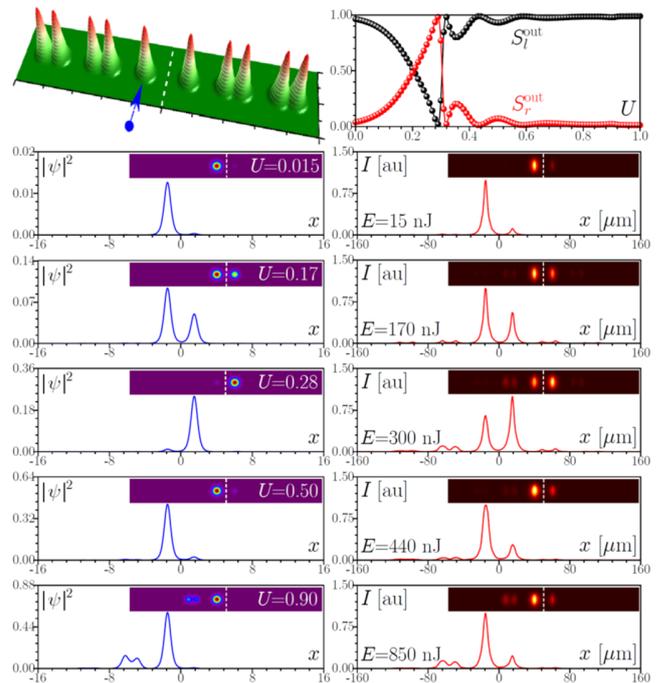

Fig. 2. First row: Schematic representation of the interface between two topological arrays with indication of the excited waveguide in the left array and theoretically calculated output power sharing between left, $S_l^{\text{out}}$, and right, $S_r^{\text{out}}$, arrays as a function of input power $U$. Second to sixth rows: Comparison of the theoretical (blue lines) and experimental (red lines) output intensity cross-sections at $y = 0$ and 2D intensity distributions (insets) for increasing energies $E$. Vertical dashed lines in the insets separate two arrays and serve as guides for the eye. Here $s_l = s_r = 15$ μm, $d_l = d_r = 33$ μm, $h = 30$ μm, and $p = 4.88$.

For the experimental observation of the edge state switching we fabricated a set of TT and TN structures with spacing $h$ between two arrays ranging from 15 μm to 46 μm. In the experiments, light from 1 kHz femtosecond Ti:sapphire laser system Spitfire HP (Spectra Physics) delivering pulses with 40 fs duration with 800 nm central wavelength first passes through an active beam position stabilization system (Avesta) and an attenuator, and afterwards enters into a two-pass single-grating stretcher-compressor with a variable slit that selects the necessary spectral components to optimize the output pulse width. Maximal nonlinear localization in our arrays is observed for a spectral width of 5 nm and pulse duration of $\tau \sim 300$ fs (FWHM). Such stretched pulses were focused into the outermost waveguide in one of the arrays, as schematically indicated by the blue arrow in the top row of Figs. 2, 3 and 5, 6. Since in the TT case both arrays are equivalent, we excite the edge waveguide in the left array. Output intensity distributions were registered with 12.3 MP scientific CMOS camera Kiralux (Thorlabs). Taking into account losses for matching to the waveguide mode, the input peak power in the waveguide can be defined as the ratio of the pulse energy $E$ to pulse duration $\tau$ and is evaluated as 2.5 kW for each 1 nJ.

The observation of nonlinearity-controlled switching between two topological arrays for array spacing $h = 30$ $\mu$m is presented in Fig. 2. The excited edge waveguide in the left array is shown by blue arrow. For this spacing, in the linear case the length of the sample approximately corresponds to two beating lengths, $2L$, hence in the linear regime light first switches to the right array and then returns to the left one. To quantity the impact of nonlinearity on switching for this spacing $h$, in the top row of Fig. 2 we show the theoretically calculated from Eq. (1) dependence of the fraction of power concentrated in the left and right arrays given by

$$S_l^{\text{out}} = U^{-1} \int_{-\infty}^{0} dx \int_{-\infty}^{\infty} dy |\psi|^2,$$
$$S_r^{\text{out}} = U^{-1} \int_{0}^{\infty} dx \int_{-\infty}^{\infty} dy |\psi|^2,$$
(2)

on the total input power $U = \int_{-\infty}^{\infty} dx \int_{-\infty}^{\infty} dy |\psi|^2$ after 10 cm of propagation. As one can see, nonlinearity slows down the coupling resulting first in concentration of light in the right array, and eventually it completely arrests coupling, so that in the high-power regime light always remains in the left part of the array. Qualitatively, this type of switching dynamics still occurs due to the interference of two in-phase and out-of-phase modes nearly equally populated by the single-waveguide excitation. However, because nonlinearity changes propagation constant difference between these modes (it affects in-phase and out-of-phase modes differently) and slightly adjusts their profiles, one observes that the interference pattern at the output face of the sample (defining power fractions in two arrays) now changes with the input pulse energy, i.e. one can observe instances where power is concentrated in only one of the arrays or equally divided between them. Switching curves are sufficiently sharp, with fast variation of $S_{l,r}^{\text{out}}$ around $U \sim 0.3$. Our experiments fully confirm this switching scenario. In the right column of Fig. 2 we show the output intensity cross-sections at $y = 0$ as well as corresponding full 2D intensity distributions (insets) measured for various pulse energies $E$. In the linear regime, $E = 15$ nJ, light switches from the left to right array, and then returns to the left array. Increasing the pulse energy results in the gradual concentration of light in the right array, reaching a maximum around $E \sim 300$ nJ. It should be stressed that due to the pulsed nature of the excitation for this spacing $h$ some visible fraction of radiation remains at the output in the left array. This is because the high-intensity parts of the pulse switch to the right array, while the pulse tails still switch linearly, contributing to the spot in the left array. A related discussion of the impact of pulsed excitation on switching in nontopological dual-core fibers can be found in [85]. With further increase of the pulse energy up to $E \sim 440$ nJ light concentrates again in the left array and remains there over a wide interval of energies. Because in our system switching occurs for modes belonging to the topological gap, at high enough power the propagation constants of such modes may be driven by nonlinearity into the top allowed band [see Fig. 1(c)], facilitating the coupling with the bulk states. Such coupling is clearly seen in the last row of Fig. 2 for $E \sim 850$ nJ. The theoretical distributions obtained from Eq. (1) for the single-site input are presented in the left column of Fig. 2. Since we use a spatial model [i.e. pulsed nature of excitation is not taken into account in Eq. (1)], theory predicts a complete concentration of light in the right array around $U = 0.28$. By and large, one can conclude that main switching features, including coupling with the bulk modes at high input powers are very accurately reproduced in the theory. We also stress that the observed switching is a robust phenomenon: fluctuations of the output power fractions concentrated in two arrays remain very small at all power levels and they are connected only with small fluctuations of power in the input laser beam.

Now we consider a similar TT configuration, but with increased spacing $h = 36$ $\mu$m between the two topological arrays, see Fig. 3. Now 10 cm sample corresponds to approximately one beating length $L$ between symmetric and antisymmetric interface modes 1 and 2. As a result, light launched into the left array (see blue arrow in the top row) completely switches to the right one at the output. Switching curves in the form of dependencies of power fractions in the left and right arrays $S_{l,r}^{\text{out}}$ on the input power $U$ show sufficiently smooth growth of power in the left array. The experimental output cross-sections and 2D intensity distributions are shown in the right column of Fig. 3 and can be compared with the corresponding theoretical results in the left column. Now, with increase of the pulse energy $E$ one observes a monotonic increase of the power in the left array, equilibration of powers in the two arrays around $E \sim 175$ nJ, and nearly complete concentration of light in the left array at $E \sim 370$ nJ (in this case contribution from the pulse tails leading to small spots in the right array is much less pronounced). Finally, coupling with the bulk states occurs at the same energy levels as for smaller value of spacing $h$. It should be mentioned that power-controlled switching is observable for the spacings down to $h \sim 21$ $\mu$m although in such structures multiple complete switching events (up to 3) occur on the length of the sample and there exist a pulse energy range where small variations of $E$ cause rapid variations of the power fractions $S_{l,r}^{\text{out}}$ in the two arrays.

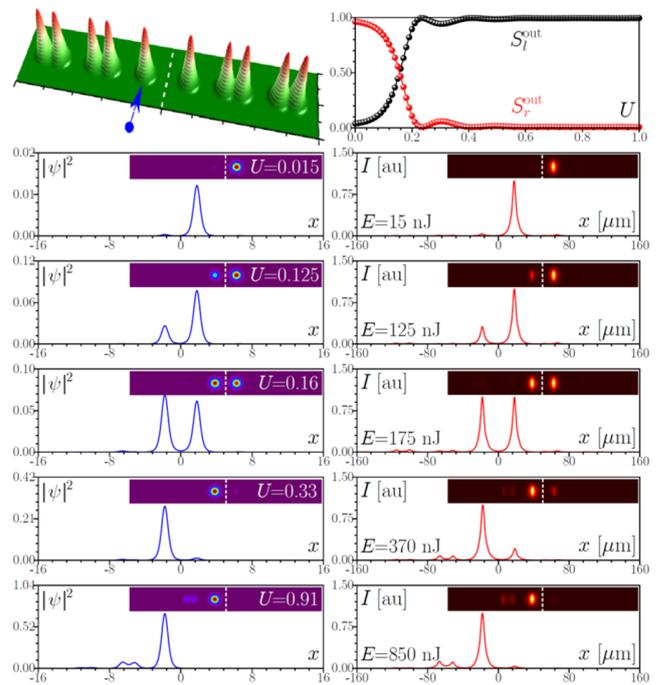

Fig. 3. Same as in Fig. 2, but for the larger spacing $h = 36$ $\mu$m between two topological arrays.

To confirm robustness of switching we also studied how disorder impacts the topological modes determining switching dynamics and their propagation constants. To do this, we considered TT arrays with small shifts of waveguide positions given by the random numbers uniformly distributed within the interval $[-\delta_s, +\delta_s]$, where $\delta_s = 0.1$. Notice that the selected $\delta_s$ value (corresponding to the shift of $1~\mu$m) exceeds the error of our positioning system used for waveguide writing by two orders of magnitude. The depths of all waveguides were also allowed to fluctuate within the range $[p(1-\delta_p), p(1+\delta_p)]$, i.e. we consider both off-diagonal and diagonal disorder. The dependence of propagation constants of modes supported by such disordered TT structure on spacing $h$ is shown in Fig. 4(a) for $\delta_p = 0.01$ and Fig. 4(b) for $\delta_p = 0.02$ (for each value of $h$ a new disorder realization was selected). One can see that the disorder leads to certain broadening of the bulk bands and fluctuations of propagation constants of the in-phase and out-of-phase modes residing at the interface between two arrays (such fluctuations always occur in continuous systems [65]). However, both these modes are clearly present in the spectrum (as long as the topological gap exists), and only for the large spacings $h$ the disorder may lead to considerable variation of switching dynamics because in this limit the fluctuations of the difference $b_1 - b_2$ of propagation constants of two modes may become comparable with this difference in the unperturbed system. In contrast, for smaller $h$ values switching dynamics will remain practically unaffected by disorder. From the point of view of switching dynamics, disorder will lead to the shifts of $z$-positions at which light fully couples to the neighboring array or returns to the input one. For linear regime with $U = 0.01$ this is illustrated in Fig. 4(c), where power sharing between two arrays is shown as a function of $z$ for different disorder realizations. The same occurs in the nonlinear regime with $U = 0.29$ (this power for selected $h = 30~\mu$m is close to the level when without disorder light couples to the right array at the output), as shown in Fig. 4(d). One can thus conclude that the reported switching phenomenon persists in the presence of disorder.

We also explored TN structures – a topological array brought into close proximity with a nontopological one. The spectrum of the modes in such combined structure is presented in Fig. 1(d). It indicates that the evolution dynamics is substantially different for excitations in the topological or nontopological arrays. In Fig. 5 we present experimental and theoretical results for the excitation of the outermost waveguide in the topological array, as shown by the blue arrow in the top row, for sufficiently large spacing $h = 31~\mu$m. This type of excitation features a large overlap only with the single localized interface mode 3 [Fig. 1(e)] existing in such a structure; other modes remain practically non-excited. As a result, one observes excitation of the only nonlinear mode bifurcating from the corresponding linear interface state (it is similar to the soliton reported in [82], see right column of Fig. 5) and no switching between the two arrays takes place due to the absence of interference with other modes [see the $S_{l,r}^{\text{out}}(U)$ dependencies in the top row]. When the pulse energy reaches sufficiently high levels, one observes coupling with bulk modes in both topological and nontopological arrays that may lead to reduction of $S_l^{\text{out}}$ and certain increase of power fraction in the right array.

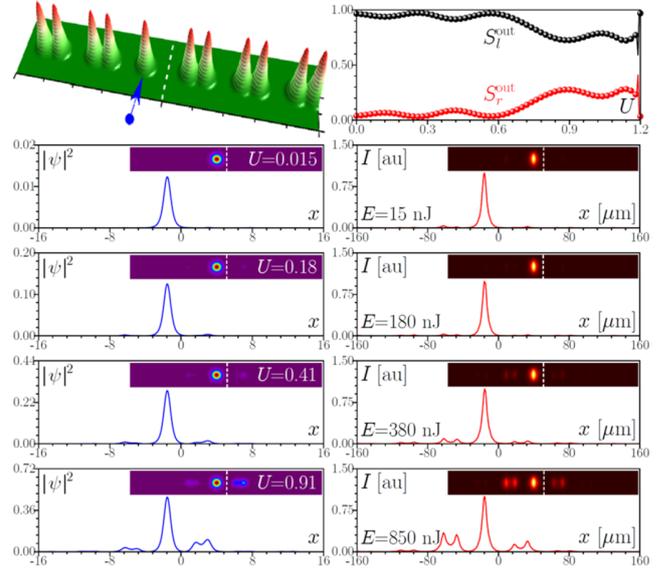

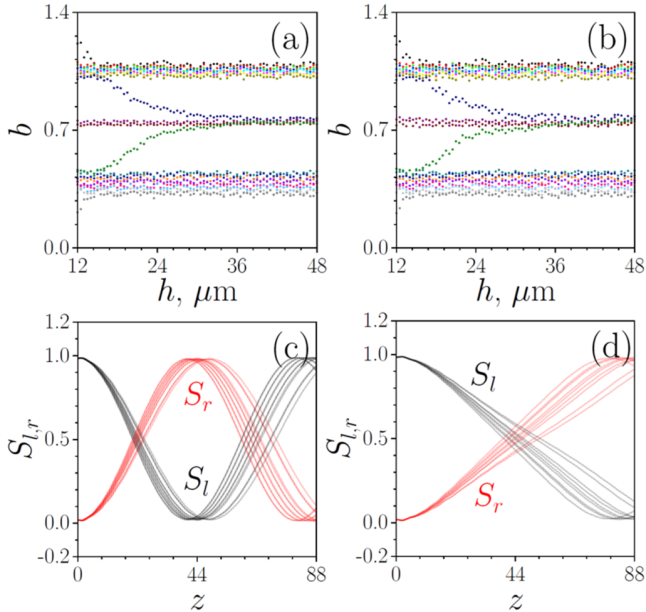

Fig. 4. Eigenvalues of the modes supported by the topological-topological ($s_l = s_r = 15~\mu$m, $d_l = d_r = 33~\mu$m) array as functions of spacing $h$ between two arrays for disorder strength $\delta_p = 0.01$, $\delta_s = 0.1$ (a) and $\delta_p = 0.02$, $\delta_s = 0.1$ (b). Power sharing between left, $S_l$, and right, $S_r$, arrays as a function of propagation distance $z$ at $U = 0.01$ (c) and $U = 0.29$ (d) for different realizations of disorder with $\delta_p = 0.01$, $\delta_s = 0.1$ and $h = 30~\mu$m. In all cases $p = 4.88$.

Fig. 5. First row: Schematic representation of the interface between the topological and nontopological arrays with indication of the excited right outermost waveguide in the topological array and output power sharing between left, $S_l^{\text{out}}$, and right, $S_r^{\text{out}}$, arrays as a function of input power $U$. Second to fifth rows: Comparison of the theoretical (blue lines) and experimental (red lines) output intensity cross-sections at $y = 0$ and 2D intensity distributions (insets) for increasing energies $E$. Here $s_l = d_r = 15~\mu$m, $s_r = d_l = 33~\mu$m, $h = 31~\mu$m, and $p = 4.25$.

Finally, our experiments confirmed that when a waveguide is excited in the nontopological array in the TN structure, one does not observe switching between the two parts of the structure. For this type of excitation at $h = 31~\mu$m most of the power remains in the nontopological part for all pulse energies, see Fig. 6. Due to the absence of localized states in this part of the structure, in the linear regime at $E \sim 15$ nJ one observes strong diffraction into the nontopological array and practically no light couples into the topological part (see the experimental distributions in the right column). At sufficiently high pulse energies of the order of $E \sim 400$ nJ, light shrinks to a pair of close waveguides at the edge of the nontopological array and undergoes periodic oscillations between these two waveguides.

In conclusion, we have reported the experimental observation of switching between topological edge states with propagation constants within the topological bandgap of the supporting waveguide arrays. Switching occurs only between two topological structures and it can be controlled by the input pulse energy. Our observations motivate the exploration of the impact of nonlinearity in the behavior of general topological switching devices and, in particular, can be readily extended to photonic valley Hall and Floquet topological insulator schemes based on waveguide arrays.

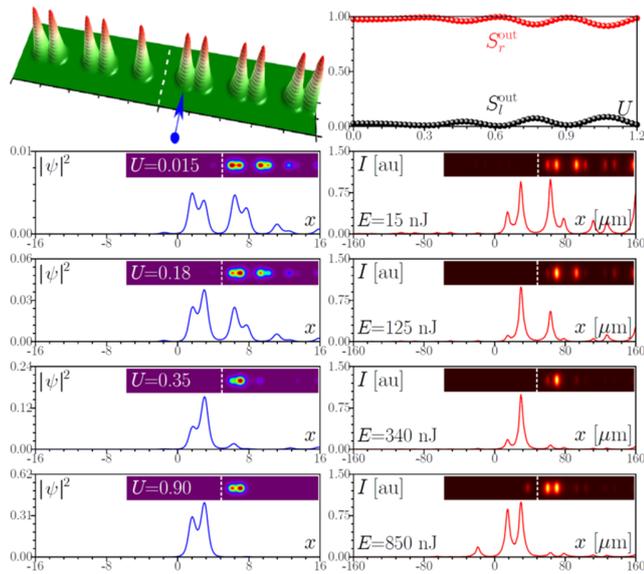

Fig. 6. Same as in Fig. 5, but for the excitation of the left outermost waveguide in the nontopological array.

**Acknowledgements:** The authors acknowledge funding of this study by RSF (grant 21-12-00096).
**Author contribution:** All the authors have accepted responsibility for the entire content of this submitted manuscript and approved submission.
**Conflict of interest statement:** The authors declare no conflicts of interest regarding this article.